\documentclass[usenatbib,usegraphicx]{mn2e}
\usepackage{times}
\newcommand{\kms}{km.s$^{-1}$ }

\begin{document}
\title{VLBA imaging of a periodic 12.2 GHz methanol maser flare in  G9.62+0.20E  }

\author[Goedhart, Minier, Gaylard \& van der Walt]
{S. Goedhart$^{1,2}$\thanks{E-mail: sharmila@hartrao.ac.za}, V. Minier${^3}$, M. J. Gaylard$^{1}$ and
D. J. van der Walt$^{2}$ \\ 
$^1$ Hartebeesthoek Radio Astronomy Observatory, PO Box 443,
Krugersdorp, 1740, South Africa\\ 
$^2$     School of Physics,
North-West University, Potchefstroom campus, Private Bag X6001, Potchefstroom, 2520, South Africa \\
$^3$ Service d'Astrophysique, DAPNIA/DSM/CEA Centre d'Etudes de Saclay, 91 191 Gif-sur-Yvette, France
}

\date{Accepted ;      Received ;      in original form }

\pagerange{\pageref{firstpage}--\pageref{lastpage}}
\pubyear{2004}

\maketitle

\begin{abstract}
The class II methanol maser source G9.62+0.20E undergoes periodic flares at both 6.7 and 12.2 GHz.  The flare starting in 2001 October was observed at seven epochs over three months using the VLBA at 12.2 GHz. High angular resolution images (beam size $\sim$ 1.7 x 0.6 mas) were obtained, enabling us to observe changes in 16 individual maser components. It was found that while existing maser spots increased in flux density, no new spots developed and no changes in morphology were observed. This rules out any mechanism which disturbs the masing region itself, implying that the flares are caused by a change in either the seed or pump photon levels. A time delay of 1--2 weeks was  observed between groups of maser features. These delays can be explained by light travel time between maser groups.  The regularity of the flares can possibly be explained by a binary system.
 
\end{abstract}


\label{firstpage}

\section{Introduction}

Class II methanol masers appear to be closely associated with the earliest evolutionary stages of high mass stars \citep[][and references therein]{Men02,Min03}.  In many cases, the masers at 6.7 or 12.2 GHz are the only indication of star formation activity at a particular site \citep{Wal98,Goe02}.  Thus methanol masers are excellent probes of the changing conditions in the vicinity of young (proto)stars.

The prevalent model for class II methanol masers is the Sobolev-Deguchi model \citep{Sob94,Sob97}. This model has recently been further refined to examine in detail the large number of methanol maser transitions observed in star-forming regions \citep[e.g.][]{Cra01}. The model of \citet{Sob97}   is the only one which has been able to reproduce the high brightness temperatures ($>$10$^{12}$ K) observed toward strong methanol maser sources.  In this model, the masers arise in a spherically expanding cloud of methanol-rich gas, which is treated under the large velocity gradient (LVG) approximation. The model has approximately ten free parameters, with the most important in terms of producing strong methanol masers being the gas temperature, the dust temperature, the density, the methanol column density and the beaming of the source, which characterises the elongated maser geometry. Three external sources of continuum are taken into account: the infrared pumping radiation, the cosmic microwave background and possible free-free emission from a HII region. The model requires the temperature of the gas producing the masers to be moderately cool (30 K), with a hydrogen number density in the range of $10^6-10^8$ cm$^{-3}$. In addition a nearby region of warm (100-200 K) dust is necessary to produce the infrared photons needed to create the population inversion. A dust temperature $>100$~K plays a dual role in producing methanol masers; it is required to release methanol from the dust grains through evaporation and also to pump the methanol molecules to their first and second torsionally excited states. Finally a methanol column density of $\sim5\times10^{17}$~cm$^{-2}$ is needed to account for the observed brightness temperatures.

The region G9.62+0.20 is a high mass star formation complex with a number of HII regions in different evolutionary stages.  \citet{Gar93} labelled the HII regions A to E in order of increasing compactness.  A and B are extended HII regions, C is a bright compact HII region, D is a bright ultracompact (UC) HII region and E has been classified as a hyper-compact HII region by \citet{Kur02}. Regions C, D and E are connected by a ridge of dense molecular material with C and D to the north and south of E, respectively \citep{Hof96a}.  A sixth object F, located midway between E and D, was found to have emission from hot, dense molecular dust \citep{Hof96a}. Recent observations by \citet{Tes00} found faint thermal radio continuum emission associated with component F. This source is believed to be in the hot molecular core stage  and has a high velocity outflow associated with it  \citep{Hof01}.  

Sensitive mid-infrared imaging at 11$\mu$m  by \citet{DeB03a} found three dense cores which matched the positions of HII regions B, C and E.  The mid-infrared emission at G9.62+0.20E is very weak. No near-infrared source was found at this location \citep{Per03}. 

\citet{Fra00} calculated the spectral index of G9.62+0.20E from 2.7 mm to 3.5 cm  as $0.95 \pm  0.06$ while \citet{Hof96a} found a value of $1.1 \pm 0.3$.  \citet{Hof96a} find this to be consistent with either an ionized stellar wind or a hypercompact HII region (radius $<$ 0.0025 pc),  with excess dust emission at 2.7 mm and a central  star that is probably of spectral type B1.

The observations of \citet{Hof01} also found hot methanol at HII region E, but nowhere else in the complex.  The most  powerful 6.7 GHz methanol maser known is associated with HII region E \citep{Phi98}, while a weaker maser is associated with region D.  Other maser transitions towards region E include methanol at 12.2 GHz \citep{Cas95b}, 37.7 GHz \citep{Has89}, 85.5 GHz \citep{Cra01} and  107 GHz \citep{Val99}; hydroxyl at 1665 and 1667 MHz \citep{Cas98}; ammonia at 24.5 GHz \citep{Hof94} and water masers in a line between E and F \citep{Hof96}.  The water masers are probably associated with the outflow from the hot molecular core F.  High-resolution observations of the 12.2 GHz methanol masers by \citet{Min02} show clusters of maser spots in two lines separated by $\sim$ 500 AU.  

A flux density monitoring programme at the Hartebeesthoek Radio Astronomy Observatory (HartRAO)  found that the methanol maser in  G9.62+0.20E exhibits strong  flares recurring with a period of 246 days at both 6.7- and 12.2 GHz \citep{Goe03,Goe04}.  This is the first reported incidence of periodic variations associated with a high mass star formation region.   The likely mechanism is strongly  constrained by the period observed.  High mass stars on the main-sequence can undergo $\beta$ Cephei pulsations, but these periods are between 0.1 -- 0.6 days.  Examination of the classes in the General Catalogue of Variable Stars \citep{GCVS} show that the only pulsating stars with the appropriate periods are Miras.  These are evolved stars and therefore unlikely to be the cause of the variations.  The rotational velocities of main-sequence O and B type stars are typically of the order of 400 \kms \citep{Lan92}, which gives rotational periods between 0.5 and 2 days.  A recent study of pre-main-sequence periodicity in an OB association found rotational periods ranging from less than a day up to 28 days \citep{Mak04}.  Again, these periods are far too short to be related to the variations in the masers.

The young star FU Orionis is the prototype of a class of YSOs that experience episodic outbursts due to high disc accretion rates and instabilities in the disc accretion.   The duration of these outbursts is of the order of decades and the interval between outbursts is believed to be  thousands of years \citep[][and references therein]{Ken00}.  The time-scales involved here are far too great to present a feasible explanation of the methanol maser flares. 

The periodic variations could be modulated by a disk-outflow system. The numerical simulations of  \citet{Yor02} showed that massive stars can form from disc accretion collapse of molecular cloud clumps. The simulations of \citet{Dur01}, \citet{Yor02} and \citet{Ouy03} showed that the accretion disc develops a rotating spiral density pattern. On the other hand, the outflows themselves could cause variations with the appropriate periods;  \citet{Ouy03} found that the outflow jets can develop a corkscrew structure or a wobbling motion.  The rotation rate of the corkscrew is dependent on its distance from the origin, with the minimum rotation rate at approximately 85 days for a B0 star.  G9.62+0.20E is not directly associated with an outflow, although the core at region F does have an outflow. 

Another possibility is that the central star is a binary system.   More than 50 \% of optically visible  stars of types  O9 to M  appear to have  companions \citep{Abt79,Hal83,Abt83}. \citet{Pov82} showed that the incidence of binary and multiple systems may be closer to 100 \% once observational limitations have been taken into account.  A recent study of the multiplicity in Herbig-Haro sources gives a binary frequency between 79\% and 86\%, with half  of the systems being higher order multiples \citep{Rei00}.  Simulations of accretion and collisions in stellar clusters \citep{Bon02} indicate that binary systems can form through three-body interactions. In addition, it is known that high mass stars tend to form in clusters \citep{Cla00}. Therefore there is a strong possibility that the central star exciting the masers could have a companion.  Assuming a 11 M$_\odot$ star (B1) with a less massive companion, the orbital radius of the companion would be 1.7 AU to give a period of 246 days.  

Thus the presence of periodic variations in these masers  could lead to important insights into the high-mass star formation process.

Single-dish monitoring at HartRAO shows   time delays occur in the flaring of different maser features \citep{Goe03} but the velocity overlap of some of the maser spots means that the spectral line observations cannot give us much information on the spatial propagation of the flare.  Imaging of the masers during a flare would define the projected spatial sequence of the flaring and thus help to identify the mechanism causing the flare, or eliminate some possibilities.  Since this  source is reasonably strong at 12.2 GHz and the flares are even more pronounced than at 6.7 GHz, the VLBA, with its 12 GHz capability, was judged to be the ideal instrument for imaging the maser components.  Seven observations were done during the period of October to December 2001.

The observations and reduction method are described in Section 2.  The results are presented in Section 3 and discussed in Section 4.  The conclusions are presented 
in Section 5.

\section{Observations and reduction }
The observation dates  were selected such that the various phases of the flare could be covered effectively. There is a period of about two weeks during which the main peak of the maser slowly increases in intensity.  It was judged that one observation would be sufficient during this period.  The maser then  rapidly increases in intensity over the next two weeks and remains at its peak for less than a week before starting to decay.  Four observations at weekly intervals were requested during this stage. The decay phase takes another two months, but just two more observations at two-week intervals were used.  

\begin{figure}
\resizebox{\hsize}{!}{\includegraphics[clip,angle=0]{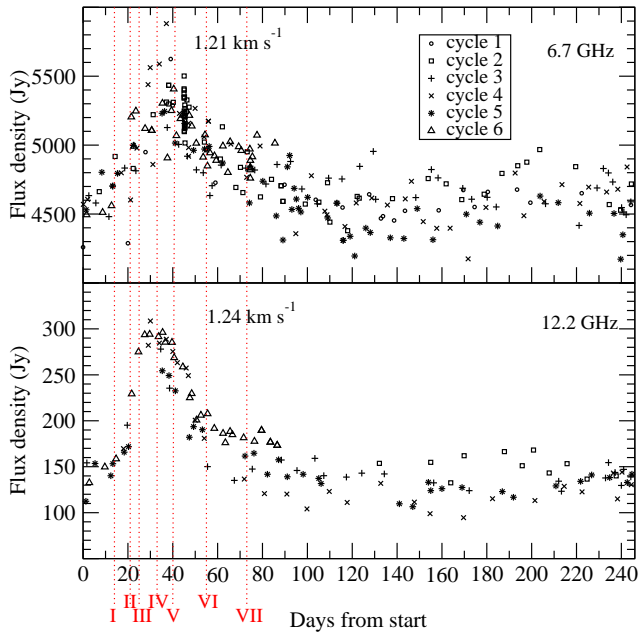}}
\caption[Timing of VLBA observations relative to the flare cycle of feature C]{Timing 
of the VLBA observations relative to the flare cycle of feature C in G9.62+0.20E.  The red lines indicate the times of the VLBA 
observations during cycle 5.}
\label{fig:vlba-cycles}
\end{figure}

Table~\ref{tab:dates} gives the dates of the observing runs, along with an estimate of the rate of the flare from the folded time-series of several cycles. Figure~\ref{fig:vlba-cycles} shows the timing of the observations relative to the flare cycles. The progress of the flare during this time was  closely monitored at HartRAO -- as described in \citet{Goe03} -- in order to ensure that the VLBA observations were correctly timed\footnote{We would like to thank Mark Claussen at the NRAO for making sure that the observations were correctly scheduled and for preparing the \textsl{SCHED} file.}.

The observations, at both left and right circular polarisation, were carried out using the full VLBA at 2cm.  The continuum source J1733-1304 was used as a calibrator.  The sequence of scans consisted of  5 minutes on the calibrator and 30 minutes on-source.  The total observing time per session, including the calibration observations, was 6 hours. Since the maser features span $\sim$ 10 km.s$^{-1}$,  a bandwidth of 1 MHz (covering 25 km.s$^{-1}$) was used.  The observations were correlated at the VLBA correlator in Soccoro (USA) using 512 channels, giving a velocity resolution of 0.05 km.s$^{-1}$.   The central velocity (at channel 257)  was 1.28 \kms and a rest frequency of 12178.595 MHz was adopted.

\begin{table}
\begin{center}
\caption{Timing of the VLBA observations}
\label{tab:dates}
\begin{tabular}{lrrrr}
\hline
Observation & Date &  Day of cycle & rate of flare\\
                    &         &                      &      Jy/hour\\
\hline
I & 11-10-2001 & 14  & 0.19\\
II & 18-10-2001 & 21 & 0.13\\
III & 22-10-2001 & 25 & 0.07\\
IV & 30-10-2001 & 33 & 0.06\\
V & 07-11-2001 & 41  & 0.14\\
VI & 21-11-2001 & 55 & 0.06\\
VII & 09-12-2001 & 73 & 0.01\\
\hline
\end{tabular}
\end{center}
\end{table}

\begin{figure*}
\resizebox{\hsize}{!}{\includegraphics[clip,angle=0]{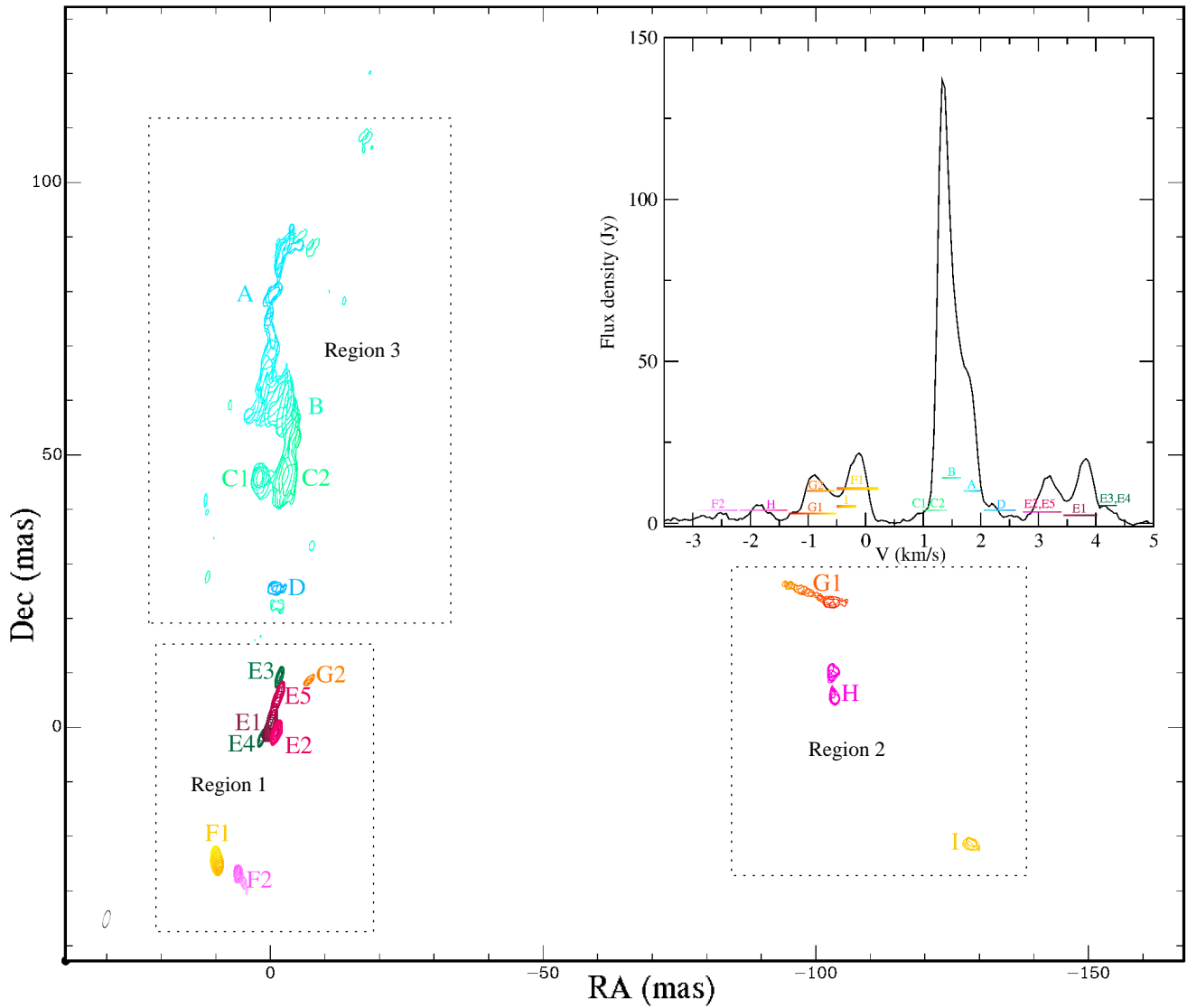}}
\caption[Distribution of maser features in G9.62+0.20E]{Distribution of maser features in G9.62+0.20E. The figure was taken from the fourth VLBA observation, when features A, B, C and D were at their maximum intensity. The features are colour-coded according to velocity.  The contour levels are at 0.2 Jy/beam in each velocity channel.  The inset shows the 12.2 GHz total power spectrum.  The colour-bars superimposed on the spectrum indicate the corresponding colour range in the contour map and the features are labelled on the contour map and on the spectrum. Some spurious features due to the sidelobes of the beam can be seen at the same velocity as features C1 and C2.  The restoring beam is shown to the bottom left.}
\label{fig:spots} 
\end{figure*}

The data were calibrated and reduced using standard procedures in the Astronomical Image Processing System (\textsl{AIPS}). 

Correlator digitisation errors were corrected using the task \textsl{ACCOR}. The data were examined for bad scans and flagged using \textsl{SPFLAG} and \textsl{IBLED}.  The fringe rate for the continuum calibrator was found using \textsl{FRING}, before using \textsl{BPASS} to find the time-dependent bandpass corrections for both the auto- and cross-correlated spectra.  The velocity information for the source was entered into the catalogues using \textsl{SETJY}.  The Doppler changes in velocity were corrected using \textsl{CVEL}.  This task also applies the bandpass calibrations to the data.

There are two options for the amplitude calibration.  The first is to use an external  continuum calibrator observed at regular intervals.  This has the disadvantage of requiring more observing time off the source of interest.  The other option is to select a total-power spectrum from a reliable antenna as a template spectrum, and use the gain calibration for that antenna.  We adopted this  method  and assumed that the amplitude of the maser stayed constant during the six hours of the observation.  The maximum rate of change in the intensity of the dominant maser feature is 0.2 Jy/hour, which gives a change of 1.2 Jy, at most, over 6 hours.  The template spectrum had to have no bad data in it and be taken at maximum elevation.  The most important criterion was to make sure that there were no signs of RFI during the scan since this leads to errors in the recorded system temperature.  The system temperatures for the scan were examined carefully to make sure that a constant value was recorded.  Once a suitable template spectrum was found, it was compared to all the scans at each antenna and a gain correction calculated relative to it, using the task \textsl{ACFIT}.  The \textsl{ACFIT} solution was examined and note was taken of solutions with extremely high gain, indicating high attenuation of the signal, usually due to dense cloud, rain or the source being low on the horizon.  Once the amplitude calibration was applied, the data with high gain corrections were flagged using \textsl{IBLED}.

The time-dependent phase delay, or instrumental delay, was found by running \textsl{FRING} on the continuum calibrator.  The delay solution was smoothed using \textsl{SNSMO} and then applied to the maser source using two-point interpolation.  

The phase rate calibration was found using a compact maser feature as the source model in \textsl{FRING}.  The channel used corresponded to a feature at 3.6 \kms and was visible in the cross-power spectra for the longest baselines, indicating that it is a very compact source. The solutions were examined to make sure  they  did not  have any abrupt changes which would indicate the solution interval was too long. A solution interval of 0.5 minutes was used, while the time interval in the calibration table (CL table) was 0.1 minutes.  

The quality of the image can be improved by doing self-calibration to remove the effects of rapidly-changing atmospheric conditions  which have not been removed in the previous calibration steps. Since the science goal involved tracking changes in maser amplitude, it was important to retain the absolute amplitude calibration.  Hence no further amplitude calibrations were done, since self-calibration, while it improves the S/N, destroys the absolute amplitude calibration.   The self-calibration on the phase was done by using a compact maser spot as a reference channel.  A preliminary image using 100 \textsl{CLEAN} components from \textsl{IMAGR} was used as the source model, which was then input back into \textsl{CALIB}.  The subsequent solution was applied to the reference channel and a new image and source model was generated by \textsl{IMAGR}.  The procedure was iterated until the solution converged.  The images  were checked at each stage to ensure there was indeed an improvement in the image quality.  The rms noise of the images was checked using \textsl{IMSTAT}.  The solutions typically converged after five iterations.   The derived corrections were then applied to all the other channels. The final data cube was produced using \textsl{IMAGR} with a cell size of 0.2 mas.  The data were \textsl{CLEAN}ed until the residual flux limit was 60 mJy (about 4 times the theoretical rms noise). This self-calibration method destroys the absolute position information, hence all maps displayed in this paper are relative to the reference feature.  The nominal position of this feature is at
18:06:14.655 -20:31:31.60 (J2000), using the observations of \citet{Min02} as a reference for the map centre.

Care was taken in the identification of maser components on the final image since errors in the calibration or imaging process may lead to artifacts.  Phase errors can cause multiple identical features scattered around the correct position. The phase calibration was repeated and data with bad solutions were flagged out until no such artifacts were seen in the images.   Imperfect subtraction of the dirty beam's sidelobes can create features resembling the sidelobes. The sidelobe artifacts have an amplitude of the order of 0.5\% of the peak flux density of the maser spot, so they are not dominant in the images.

\section{Results}

\subsection{Structure of the maser components}

\begin{figure*}
\resizebox{\hsize}{!}{\includegraphics[clip,angle=0]{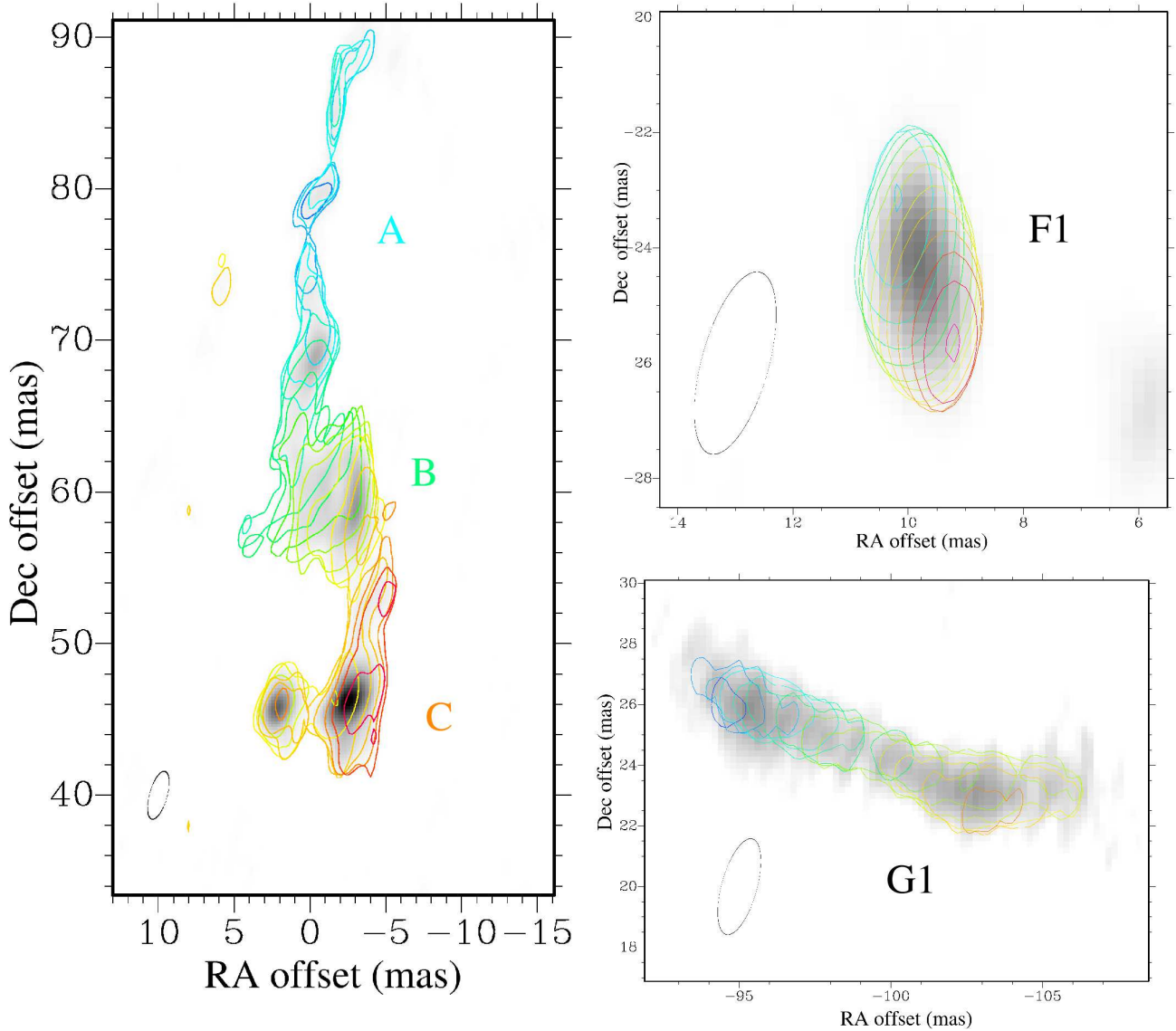}}
\caption[Close-up images of the velocity structure of selected maser components]{Close-up images of the velocity structure of various maser components. Each colour contour corresponds to a different velocity channel, with the contour limit  at 0.2 Jy/beam. The complex of components A, B and C is shown in the left-hand panel.  Component F1 is shown to the upper right and G1 to the bottom left.}
\label{fig:velocity}
\end{figure*}

\begin{table}
\begin{center}
\caption{Velocity ranges of 12.2 GHz maser components in G9.62+0.20E}
\label{tab:spots}
\footnotesize
\begin{tabular}{lrr}
\hline
Label & Start velocity & End Velocity \\
           &  \kms & \kms \\
\hline
E3 & 4.068 & 4.357 \\
E4 & 4.068 & 4.357 \\
E1 & 3.444 & 4.021 \\
E2 & 2.722 & 3.396 \\
E5 & 2.722 & 3.396 \\
D  & 2.049 & 2.674 \\
A & 1.713 & 2.001 \\
B  & 1.328 & 1.664 \\
C1 & 0.991 & 1.424 \\
C2 & 0.991 & 1.424 \\
F1 & --0.499 & 0.222 \\
I & --0.499 & --0.162  \\
G1 & --1.316 & --0.547 \\
G2 & --1.028 & --0.547 \\
H  & --2.181 & --1.364 \\
F2 &  --2.999 & --2.230 \\
\hline
\end{tabular}
\end{center}
\end{table}

The data cubes were examined using the \textsl{KARMA} \citep{Goo97} data visualisation tools \textsl{KRENZO}, \textsl{KVIS} and \textsl{XRAY}.  Of the 512 velocity channels, only the central 158 channels contain the maser emission.  The rest of the channels were disregarded in the analysis.  Sixteen individual maser components having distinct spatial positions and velocity ranges were identified. Zero-moment maps, in which the data cube is integrated along the velocity axis for specific velocity ranges, were made using \textsl{KRENZO}.  The images of \citet{Min02} show the same overall structure as seen here, but the longer integration times used here (six hours compared to about one hour) give a better \textit{uv}-coverage, with the result that feature E is resolved into five separate, compact components. Table~\ref{tab:spots} gives the 
selected velocity ranges for the components and Figure~\ref{fig:spots} shows their spatial distribution.  The nomenclature of \citet{Min02} is adopted and expanded on where necessary.  

The maser components A, B and C are connected by weaker emission. \citet{Min02} speculated that component D may be linked to components A, B and C by diffuse emission, but no evidence of such emission is seen here.  The region E has a complex structure, with five distinct compact maser components.  A channel from the component E2 was used as the reference channel for the phase calibration because of its compactness.  G1 has a linear structure with a strong velocity gradient.  H consists of two smaller 
components connected by slightly weaker emission, but it has been treated as one structure in the following analysis.  D and I are weak but compact spots.

The maser components appear to have velocity gradients.  This can be seen most clearly  in the transition from features A through C, and in F1 and G1 (Figure~\ref{fig:velocity}).  The velocity gradient in the group of components A through C is $\sim$ 0.004 km.s$^{-1}$.mas$^{-1}$, while it is 0.20 and 0.22 km.s$^{-1}$.mas$^{-1}$ in components F1 and G1, respectively.  Assuming that the distance to G9.62+0.20 is 5.7 kpc \citep{Hof96}, the velocity gradient for F1 and G1 is $\sim$ 0.03 km.s$^{-1}$.AU$^{-1}$.

\begin{figure}
\resizebox{\hsize}{!}{\includegraphics[clip,angle=0]{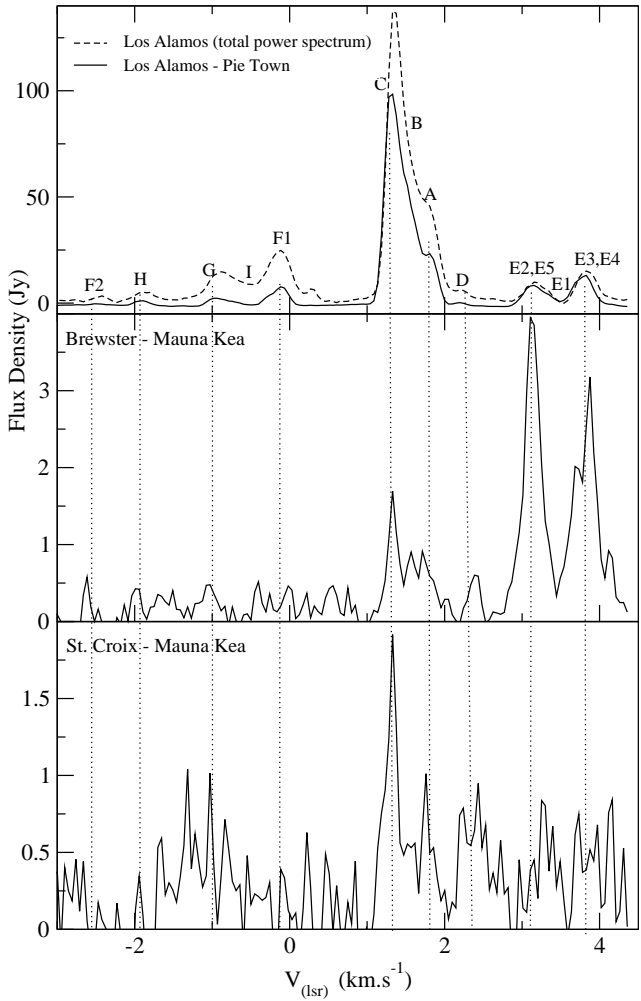}}
\caption[Examples of total power and cross-power spectra]{Examples of total power (top panel) and cross-power spectra}
\label{fig:baselines}
\end{figure}

\begin{figure}
\resizebox{\hsize}{!}{\includegraphics[clip,angle=0]{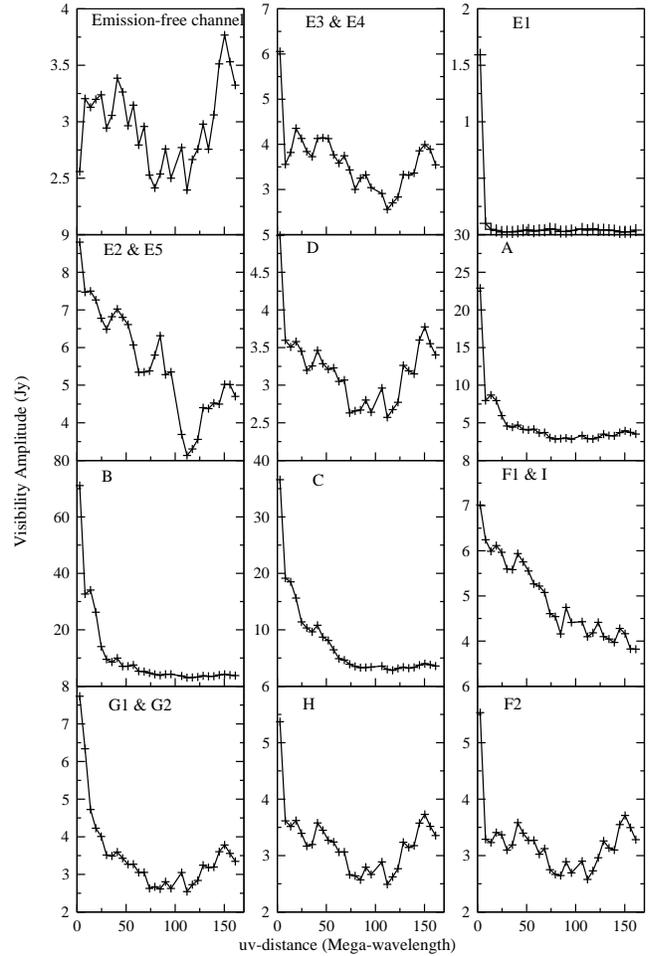}}
\caption[Visibility maplitude vs. \textit{uv}-distance for the maser features]{Visibility vs. \textit{uv}-distance for the maser features.  The visibility amplitudes have 
been averaged and  binned.}
\label{fig:uvplt}
\end{figure}

Different baselines act as filters for different scales of structures.   It is possible that some flux  could be resolved out for the features with large scale structures since there is no information from extremely short baselines.  The amount of flux that is resolved out in the cross-power spectra has to be checked as this could affect the interpretation of the variability in the images. The top panel in Figure~\ref{fig:baselines} shows a total power spectrum, which is equivalent to a single-dish spectrum, at the Los Alamos antenna and a  cross-power spectrum between Los Alamos and Pie Town, the shortest baseline in the VLBA.  Most of the flux for feature F2 has been resolved out, while there is considerable flux loss for features H, G, I, F1 and D.  There is moderate flux loss on features A, B and C while features E1--E5 do not appear to have significant diffuse emission.  The second panel shows a cross-power spectrum between Brewster and Mauna Kea, in which the dominant maser feature at 1.2 \kms is mostly resolved out, while the more compact features at 3 and 4 \kms (features E1-E5) are still clearly visible. The third panel shows the longest baseline, between St. Croix and Mauna Kea, where only the most compact features are visible in the spectrum.  It can be seen that features A, C, D, and G have compact components.  Therefore these observations will not provide information on the behaviour of the entire maser region. However, as will be discussed in the next section, the flare is still observed with the full VLBA. We assume, on this basis, that the behaviour of the compact cores can be taken as being representative of what is happening to any particular feature but it will not be possible to do any in-depth quantitative analysis such as estimating the increase in the pump or seed photons. 

Another way of examining the structure is with visibility amplitude vs. \textit{uv}-distance plots, which show the amount of emission seen at the different scales.   Figure~\ref{fig:uvplt} shows the visibility amplitudes vs. \textit{uv}-distance of the channels corresponding to each spectral feature. The top-left panel shows an emission-free channel as an estimate of the noise.  The noise will be reduced as the square-root of the number of channels averaged for each feature.  The longer \textit{uv}-distances show the smaller scale structures.  Thus greater emission at smaller \textit{uv}-distance and a rapid drop to zero in amplitude towards greater \textit{uv}-distance, indicates a source that has predominantly large scale structure and is ``resolved out''  on the longest baselines.   It can be seen that the emission from features A, B,  C and E1 comes predominantly from large scale structures but they do have  compact cores visible at the longest baselines.  The emission does not drop to the noise level for features A, B and C, implying that they may have compact cores which are not fully resolved.  The emission at longer baselines drops to the noise level for features D, G1, G2, H and F2, hence they are fully resolved at the longest baselines. These results are consistent with those of \citet{Min02}. The other features have the same velocities so it is not possible to extract the information about their structure from the amplitude vs. \textit{uv}-distance plots.

\subsection{Variations during the flare}

The antennas at Pie Town, Hancock and Fort Davis were either unavailable, or data from them was not usable because of bad weather during some of the observations.   As a result of this, the beam sizes for the different epochs are not consistent if all available antennas for a particular observation are used.  This affects the observed flux density and apparent structure of the masers. Therefore these antennas  were flagged out of all the data sets before running \textsl{IMAGR}  to produce the data sets for the following analysis.  The resulting beam parameters are listed in Table~\ref{tab:beam}.  There are still some small  variations in beam size and orientation because of temporary problems with some antennas but these should have only a minor effect on the following calculations.
\begin{table}
\begin{center}
\caption{Beam parameters for the different epochs with Pie Town, Hancock and Fort Davis flagged out for all epochs.}
\label{tab:beam}
\begin{tabular}{lrrr}
\hline
Epoch & Maj. Axis & Min. Axis & Pos Angle \\
            &  (mas) & (mas) & ($^\circ)$\\
\hline
I        &       1.7     &       0.59    &       -16.27  \\
II       &       1.8     &       0.62    &       -21.04  \\
III      &       1.7     &       0.56    &       -6.86   \\
IV      &       1.7     &       0.56    &       -14.64  \\
V       &       1.7     &       0.60    &       -14.45 \\
VI      &       1.7     &       0.69    &       -20.79  \\
VII     &       1.7     &       0.72    &       -21.90   \\
 \hline
\end{tabular}
\end{center}
\end{table}

\begin{figure*}
\resizebox{\hsize}{!}{\includegraphics[clip,angle=90]{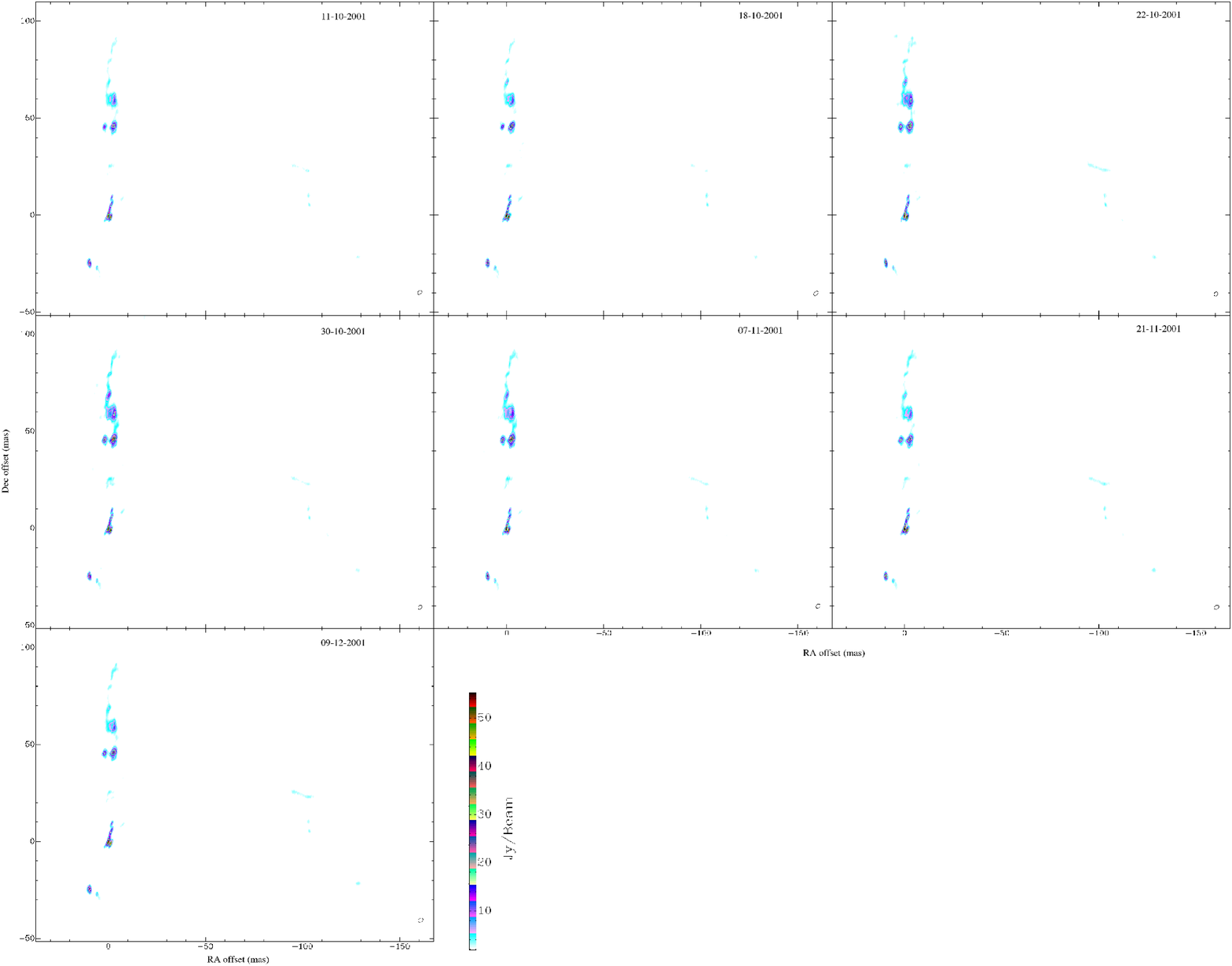}}
\caption[Zero-moment images for each epoch]{Zero-moment images for each epoch.  The restoring beam is shown to the bottom right in each panel.  An animated version of this flare sequence will be available on-line at ...}
\label{fig:movie}
\end{figure*}

Imaging of individual channels results in a ``data cube''   with an image of each channel.  Zero-moment maps were created by averaging across all the velocity channels in order to produce a single image.  The zero-moment maps for each epoch, derived from the full velocity range in which emission is present, are shown in Figure~\ref{fig:movie}\footnote{A high resolution image will be made available online}.   There appear to be no changes in the morphology of the maser components during the flare. The proper motion studies of \citet{Min01} indicate that the greatest proper motion in this source is of the order of 0.7 mas/yr and indeed, no motion of the maser features relative to the phase centre, to within 0.2 mas, was seen during the two months of the monitoring campaign. In addition, no new maser spots are formed during the flare.  The maser components simply get brighter and then start to fade  to their pre-flare intensity.

Time-series were found for each maser feature using the task \textsl{BLSUM} in the following manner.  Zero-moment maps were constructed for each of the velocity ranges given in Table~\ref{tab:spots}.  The noise levels for each map were estimated using \textsl{IMSTAT} on an emission-free region of the image (the top right-hand corner does not have emission at any velocity). The noise pixels were then set to an \textsl{INDEF} value using \textsl{BLANK}.  \textsl{BLSUM} requires a template blotch map which is used to manually select the area over which to sum.  The blotch templates were constructed by averaging the noise-blanked maps for all the epochs using \textsl{SUMIM}.  The input data cube for \textsl{BLSUM} consisted of the zero-moment maps for each epoch, so that the third axis of the cube was in fact a time axis.  \textsl{BLSUM} cannot automatically correct for the beam size in such a situation, so the flux density is given as an average brightness in Jy/beam.  An absolute flux density, which is 
independent of the beam size, can be calculated by finding the beam size in pixels/beam for each epoch, converting the average brightness to units of Jy/pixel and then multiplying by the total number of pixels in the blotch image.

\begin{figure}
\resizebox{\hsize}{!}{\includegraphics[clip,angle=0]{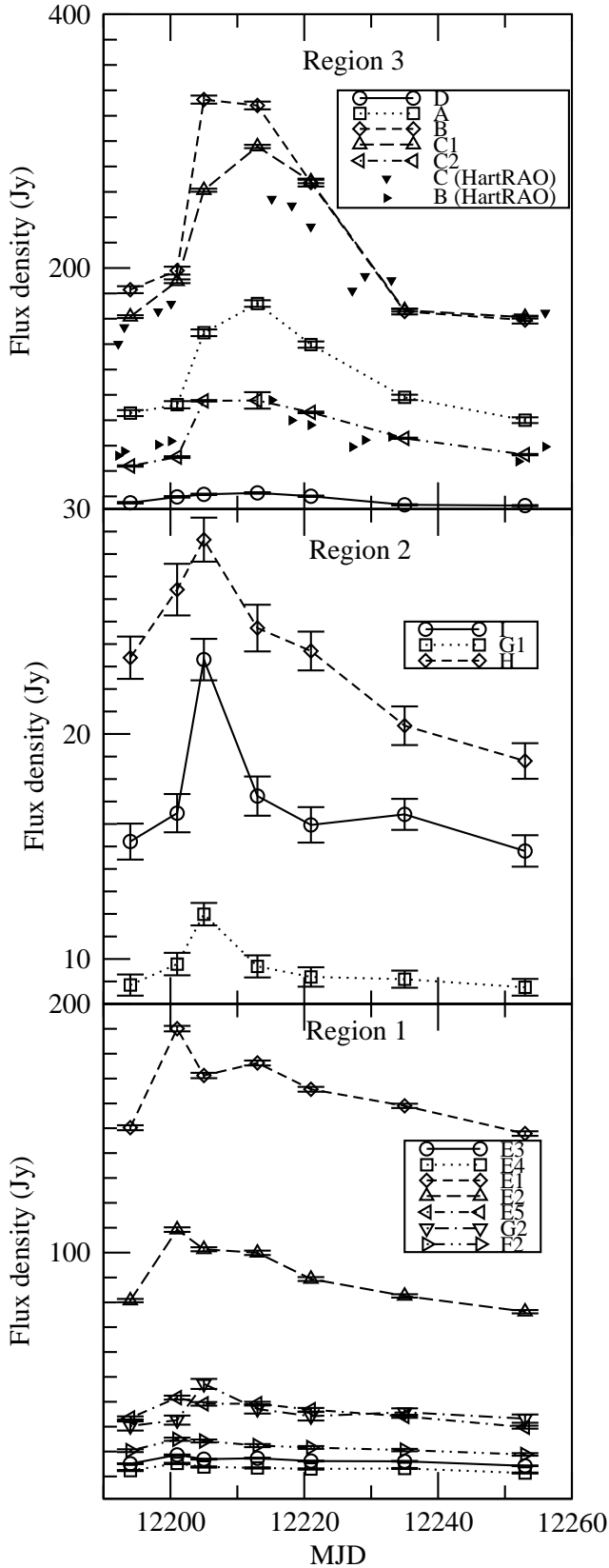}}
\caption[VLBA time-series for maser features in G9.62+0.20E]{VLBA time-series for maser features in G9.62+0.20E.  The three panels are divided according to the spatial 
distribution of the masers.  The single-dish observations at HartRAO are also shown for Features C and B.}
\label{fig:flux}
\end{figure}

The results of the calculations are shown in Figure~\ref{fig:flux}. The error bars indicate the rms noise found in each image. The HartRAO time-series of single spectral channels at the peaks of features B and C are shown in the top panel.  The maser features  A, B, C and D appear to follow the same trend as seen at HartRAO, providing an independant confirmation of the amplitude calibration of the VLBA data.  There appear to be three different trends in the time-series, depending on the positions of the masers. The masers have been grouped into three different regions based on this behaviour. The progress of the flare is best sampled amongst the features A, B, C and D.The flare appears to peak earlier at the other two groups.  The exact times of the peak of the flares at the different maser components cannot be determined because of the poor time resolution, but there may be a 1--2 week delay in the propagation of the flare through the different regions.

The flare reached its maximum between epochs I and III in region 1, while the maximum could be anywhere between epochs II and IV in region 2.  There is a gap in the observations at HartRAO due to bad weather, but from comparison with other cycles it is likely that the flare in region 3 peaked a few days before epoch IV.  Unfortunately the  S/N of the 12.2 GHz HartRAO observations of the weaker spectral features was inadequate to permit an analysis of their time-series. 

\begin{figure}
\resizebox{\hsize}{!}{\includegraphics[clip,angle=0]{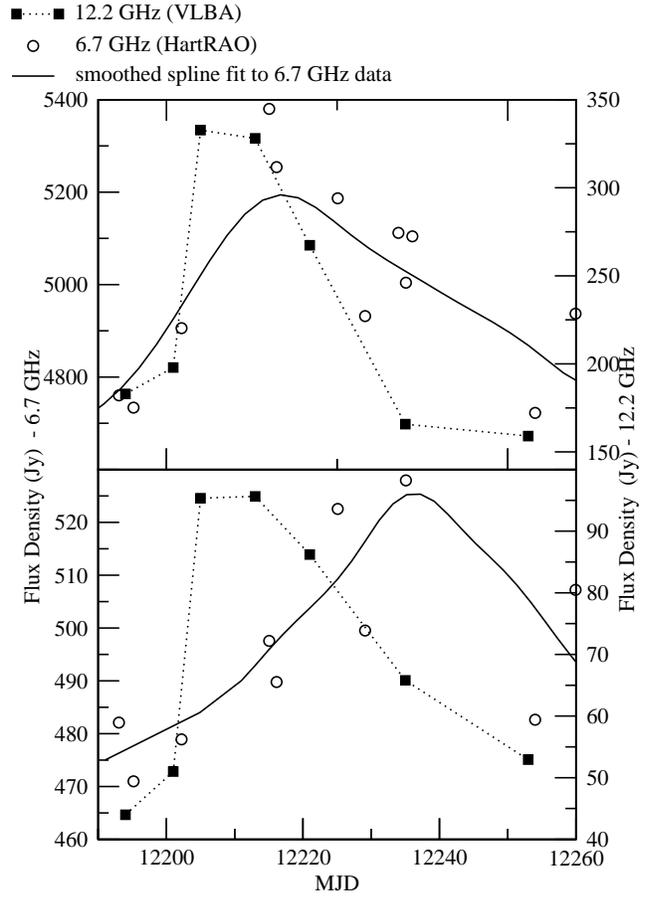}}
\caption[Comparison of flux density at 12.2 and 6.7 GHz for two velocity features]{ Comparison of flux density at 12.2 and 6.7 GHz for the 1.21 \kms feature (top panel) and the -0.14 \kms feature (bottom panel).  Smoothed spline curves (solid lines) were generated for the noisy 6.7 GHz data.}
\label{fig:compare}
\end{figure}

The maser feature at 1.21 \kms flares simultaneously at both frequencies but the -0.14 \kms feature (F1) at 12.2 GHz appears to flare approximately 30 days before the same feature at 6.7 GHz (Figure~\ref{fig:compare}). The disparity in the apparent times of the flares may possibly be explained if the 6.7 GHz maser spot is not in the same spatial region as the 12.2 GHz maser or because the pumping conditions in the two transitions are not identical.  This issue cannot be explored any further until high-resolution spot maps at 6.7 GHz are obtained and observations with better time resolution are obtained for the 12.2 GHz flares.

\section{Discussion}

\subsection{Possible flare mechanisms}

The lack of change in the structure of the masers themselves, and the fact that the masers return to almost the same pre-flare level after each flare, strongly suggests that the cause of the flare is external to the masing region. This rules out the possibility of a disturbance such as a shockwave or a clump of matter actually passing through the masing region, since this would radically change the structure of the maser components.  Two plausible possibilities are an increase in the seed photons or, assuming that the masers are radiatively pumped, an increase in the pumping radiation.  This could be caused  by stellar pulsations or periodic outbursts, or by a binary system.

If  variations in the seed photons are causing the variability, the cause of the modulation would probably be intrinsic to the star since  the cloud 
is optically thin to radiation at 6.7 and 12.2 GHz.  However, as discussed in section 1, there is no known stellar pulsation mechanism that could cause variations on the time-scales observed.  A possible means of causing periodic outbursts in the star may be repeated mass dumps from the accretion disc, similar to those seen in cataclysmic variables.  This would still require a mechanism to modulate the accretion flow rate in a periodic manner.

A close binary system seems to  be the most plausible  mechanism because of the well defined periodicity of the flares and the small orbital radius implied by the period.     As the flares last for 3 months, the modulation cannot be a simple effect, since the angular size of the stars would be too small to give rise to an effect that lasts for a third of the orbital period.  It is more likely that the effect is due to the motion of  a large condensation. It should be possible, with detailed modelling, to determine some of the characteristics of this structure since the flares are asymmetric in this source, showing a rapid rise and slow decay. Binary systems in the earliest stages of stellar evolution have not been studied at all in the case of high mass stars, but the same sort of dynamics found in low mass systems would probably apply. The environment of a young binary system can be highly complex, with interacting circumstellar discs around each star, as well as a circumbinary disc around the entire system \citep[][and references therein]{Lub00}.  In addition to this there is the surrounding molecular cloud, with possible accretion flows onto the discs.  The interaction of all of these elements may cause strong modulations of the infrared radiation reaching the maser regions.  The effects of the propagation of radiation in a region with such complex morphology cannot be easily predicted.  The optical depth of the dust is wavelength dependent and different optical depths will shift the radiation to different wavelengths.  It is not clear whether an enhancement of the radiation from the central star(s) would lead to a maser flare, or if an eclipse would shift the  wavelength of the radiation reaching the maser regions to the appropriate wavelength for maser pumping.  This problem can only be solved by modelling the radiative transfer in different scenarios.
 
\subsection{Time delays in the flare} 
 
The delay in the peak of the flares may possibly be explained from an analysis of  light travel times.  The projected sizes of the different regions are as follows:  The longest extent of region 1 is $\sim$ 44 mas, which at an assumed distance of 5.7 kpc gives 247 AU.  The light travel time across this distance is 1.3 days.  The longest extent of region 2 is $\sim$ 63 mas or 357 AU, which gives a light travel time of 1.9 days.  The longest extent of region 3 is $\sim$ 67 mas or 384 AU, which gives a light travel time of 2.3 days.   Since the interval between observations with the VLBA is of the order of  a week, delays of about a day would not be detected.  

The projected distances between the regions are as follows: The separation between regions 1 and 2 is approximately 665 AU or 3.6 days, between 1 and 3  approximately 437 AU or 2.4 days, and between 2 and 3  760 AU or 4.2 days.  The observed time delays are of the order of a week or more.  The discrepancy could be explained if the maser groups are not at the same distance along the line of sight.  Another possibility is if the modulating radiation is strongly beamed and moves across the sky in such a way as to pass across each region in succession. However, it is not obvious how such an effect or the observed pattern would be created.

\section{Conclusion}

The series of images taken over the course of three months show no change to the structure or relative positions of the maser components.    The mechanism causing the flare is therefore spatially separate from the masers.  The flares are probably caused by modulation of the pumping infrared radiation. The regularity of the flares is best explained by a binary system.  

The evidence of a delay in the propagation of the flare in different maser components is worth further investigation since this may help locate the origin of the cause of the flares, or be a means of estimating the relative offset of the maser groups along the line of sight.

{\bf Acknowledgements}
The authors would like to thank Athol Kemball for his extensive advice on amplitude calibration of spectral line VLBI data.

\bibliographystyle{mn2e}
\bibliography{../mn-jour,../refs-adsabs}
\label{lastpage}
\end{document}